\documentclass{sig-alternate}
\usepackage{latexsym,subfigure}
\usepackage{amsfonts}
\usepackage{amsmath,paralist}
\usepackage{times}
\usepackage{graphics,color}
\usepackage{url}
\usepackage{graphicx}
\usepackage{wrapfig}

\newtheorem{definition}{Definition}

\newcommand{\ignore}[1]{}

\newfont{\mycrnotice}{ptmr8t at 7pt}
\newfont{\myconfname}{ptmri8t at 7pt}
\let\crnotice\mycrnotice%
\let\confname\myconfname%

\toappear{\the\boilerplate\par
{\confname{\the\conf}} \the\confinfo\par \the\copyrightetc}

\begin{document}

\title{Emergent Behavior in Cybersecurity}

\permission{Permission to make digital or hard copies of part or all of this work for personal or classroom use is granted without fee provided that copies are not made or distributed for profit or commercial advantage and that copies bear this notice and the full citation on the first page. Copyrights for third-party components of this work must be honored. For all other uses, contact the Owner/Author.\\
Copyright is held by the owner/author(s).
} 
\conferenceinfo{HotSoS}{'14, April 08 - 09 2014, Raleigh, NC, USA}
\copyrightetc{ACM \the\acmcopyr}
\crdata{978-1-4503-2907-1/14/04.\\
http://dx.doi.org/10.1145/2600176.2600189}

\numberofauthors{1}

\author{Shouhuai Xu \\
\affaddr{Department of Computer Science}\\
\affaddr{University of Texas at San Antonio}\\
\email{shxu@cs.utsa.edu}
}

\maketitle

\begin{abstract}
We argue that {\em emergent behavior} is inherent to cybersecurity.
\end{abstract}




\section{Introduction}

The human-created cyberspace is a very large-scale complex system of cybersystems.
Its security properties are difficult to understand and characterize.
We attribute this difficulty to its {\em complexity},
a manifestation of which is the so-called {\em emergent behavior} in Complexity Science \cite{ErdiBook2008}.
Although there is no universally accepted definition of emergent behavior \cite{Kubik:2002:TFE:778855.778861},
the basic idea is intuitive. The simplest example of emergent behavior may be the well known ``$1+1>2$" effect.
For our purpose, it is sufficient to use the following informal definition of emergent behavior in cybersecurity domain.
\begin{definition}
\label{def:emergent-behavior}
A security property of a cybersystem
exhibits emergent behavior if the property is {\em not} possessed by the underlying lower-level components of the cybersystem.
\end{definition}

A direct consequence of emergent behavior is that at least some security properties cannot be understood by solely considering the
lower-level components; instead, we must explicitly consider the {\em interactions} between the lower-level components.
Although emergent behavior of cybersystems has been discussed from a function or constructional perspective
\cite{HintonNSPW97,GligorSecurityProtocol04},
emergent behavior in cybersecurity is not systematically examined until now.

\section{Emergent Behavior}

We demonstrate emergent behavior in cybersecurity through three examples.

\medskip

\noindent{\bf Example 1: Emergent behavior exhibited by cybersecurity dynamics.}
We refer to \cite{XuHotSOS14-cybersecurity-dynamics-poster} for an exposition of the emerging field of cybersecurity dynamics.
In order to explain how emergent behavior is exhibited by cybersecurity dynamics,
we consider the perhaps simplest model \cite{WangTISSEC08}.
For our purpose, it suffices to consider two cybersystems that respectively induce attack-defense structures (or graphs) $G_i=(V_i,E_i)$,
where $V_i$ is the node (vertex) set and $E_i$ is the edge set for $i=1,2$.
Let $\lambda_1(G)$ denote the largest eigenvalue of the adjacency matrix of a graph $G$,
$\beta$ denote the defense capability in detecting and cleaning compromised nodes (e.g., the probability that a compromised node gets cleaned at
a time step), and $\gamma$ denote the attack capability in compromising secure nodes
(e.g., the probability that this event occurs over an edge at a time step).
For simplicity, suppose $G_i$ is a complete graph with $n_i$ nodes for $i=1,2$.
It is well known that $\lambda_1(G_1)=n_1-1$
and $\lambda_1(G_2)=n_2-1$.
For $i=1,2$, if $\lambda_{1}(G_i)<\beta/\gamma$, the attacks will eventually be wiped out in  the cybersystem that induces
$G_i$; if $\lambda_{1}(G_i)>\beta/\gamma$, the attacks cannot be wiped out \cite{WangTISSEC08}.

Now we consider a new cybersystem that is obtained by interconnecting the aforementioned two cybersystems that induced $G_1$ and $G_2$.
Consider the simplest case that any node can attack any other node in the interconnected cybersystem,
which effectively induces attack-defense structure, a complete graph, $G_{1,2}$ with $n_1+n_2$ nodes and $\lambda_1(G_{1,2})=n_1+n_2-1$.
In many (if not all) cases,
the defense capability $\beta'$ and the attack capability $\gamma'$ associated to $G_{1,2}$ are respectively
the same as the defense capability $\beta$ and the attack capability $\gamma$ associated to $G_1$ and $G_2$.
Since $\lambda_{1}(G_i)<\beta/\gamma$ for $i=1,2$ do not imply $\lambda_{1}(G_{1,2})<\beta'/\gamma'=\beta/\gamma$,
we conclude that the attacks can be wiped out in the two underlying component cybersystems,
but cannot be wiped out in the interconnected cybersystem as long as $\lambda_{1}(G_{1,2})>\beta/\gamma$.
This phenomenon can be naturally extended to more sophisticated settings (e.g., \cite{XuTAAS2012,XuInternetMath2013-sub,XuTAAS2014,XuTDSC2012}).
This implies that cybersecurity dynamics cannot be determined by looking at the component cybersystems alone.
Rather, we need to look into how the component cybersystems interact with each other.

\medskip

\noindent{\bf Example 2: Emergent behavior exhibited by security properties in the extended trace-property framework.}
In the field of program verification, it was known that specifications that are sufficient for {\em sequential} programs are not sufficient for
{\em concurrent} programs.
For dealing with concurrent programs, Lamport proposed the safety-liveness framework of trace properties \cite{LamportIEEETSE77}.
Intuitively, a {\em trace} is a finite or infinite sequence of states corresponding to an execution of a program.
A {\em trace property} is a set of traces such that every trace, {\em in isolation}, satisfies the same predicate.
A {\em safety} property says that no ``bad thing" happens during the course of a program execution,
while a {\em liveness} property says that ``good thing" will eventually happen during the course of a program execution.
Both safety and liveness are trace properties.
A beautiful result is that every trace property is the intersection of a safety property and a liveness property \cite{LamportIEEETSE77,SchneiderIPL1985}.

Given the above history,
it is appealing to specify a cybersystem as a set of traces, and therefore as a subset of a security property
that is also specified as {\em a set of traces}.
Unfortunately, security properties are
not trace properties as shown in
\cite{GoguenOakland82,ClarksonJCS2010,SchneiderDHC2011} and refreshed below.
First, {\em noninterference} is a security property that captures the intuition that
system security is preserved as long as high-clearance (or high-privilege) processes cannot influence the behavior of low-clearance (low-privilege) processes.
It is no trace property because it cannot be verified without examining the other traces.
Second, {\em information-flow} captures some kind of correlation between the values of variables in multiple traces.
It is no trace property because it cannot be verified by examining each trace alone.
Third, {\em average service response time} is an availability property. It is no trace property
because it depends on the response time in all traces.

In an effort to overcome the above limitation of the safety-liveness framework,
Clarkson and Schenider extended the notion of {\em trace properties} to the notion of {\em trace hyperproperties} \cite{ClarksonJCS2010}.
Basically, hyperproperties are {\em sets of trace properties}.
In parallel to the safety-liveness framework, a hyperproperty is also the intersection of a safety hyperproperty and a liveness hyperproperty.
It is now known that information-flow, integrity and availability can be hypersafety or hyperliveness \cite{ClarksonJCS2010}.
Exactly because hyperproperties capture that the verification procedure {\em must} examine
across {\em multiple} traces, which may accommodate interactions between the component systems, we say that hyperproperties
exhibit the emergent behavior.
This means that we need to study the emergent behavior in cybersecurity,
which may explain why it took so long to realize the importance of hyperproperties.

\medskip

\noindent{\bf Example 3: Emergent behavior exhibited by cryptographic security properties.}
Cryptographic secure multiparty computation allows multiple parties $P_1,\ldots,P_m$, each having
a respective secret $x_1,\ldots,x_m$, to compute a function $f(x_1,\ldots,x_m)$ such that no information
about the $x_i$'s is leaked except for what is implied by the output of the function.
This manifests a confidentiality property.
A beautiful feasibility result is that any polynomial-time computable
function $f(\cdot,\ldots,\cdot)$ can be securely computed \cite{Yao86,GMW87},
as long as the protocol executes {\em in isolation} (the stand-alone setting) and trapdoor permutations exist.
When such cryptographic protocols are used as building-blocks in larger applications/systems,
they may execute concurrently (rather than in isolation). This leads to a natural question: Are the cryptographic protocols,
which are provably secure when executed in isolation, still secure when they are concurrently called by larger applications/systems?
Intuitively, concurrent executions offer the attacker the leverage (for example) to schedule the messages
in a way that is to the attacker's advantage, which does not have a counterpart in the stand-alone setting.

Quite similar to what happened in the field of program verification, where
specific properties (e.g., partial correctness and mutual exclusion) were investigated before the introduction of the
unifying safety-liveness framework \cite{LamportIEEETSE77}, the same kind of development was made in the field of cryptographic protocols.
That is, specific cryptographic security properties
were investigated before
the introduction of the unifying notion called {\em universal composability} \cite{CanettiFOCS01}, or its equivalent (but perhaps more intuitive)
version called {\em concurrent general composition} (arbitrarily many instances run possibly together with arbitrary other protocols) \cite{LindellFOCS03}.
It is now known that there are cryptographic multiparty computation protocols, which are provably secure when executed in isolation, but
are {\em not} secure when they are concurrently called by larger applications/systems.
For example,
there exist classes of functions that cannot be computed in the universally composably secure fashion \cite{CanettiEurocrypt03}.
In other words, these functions can be securely computed by running some cryptographic protocols in isolation,
but cannot be securely computed when the protocols execute concurrently.
In order to make cryptographic multiparty computation protocols secure when they are used as building-blocks for constructing larger cybersystems,
we need to make extra assumptions, such as that majority of the parties $P_1,\ldots,P_m$ are not compromised \cite{CanettiFOCS01}.
This manifests emergent behavior.
(It is interesting to note that whether or not it is reasonable to assume that
majority of the parties are not compromised may be addressed by the cybersecurity dynamics framework \cite{XuHotSOS14-cybersecurity-dynamics-poster}.)

\medskip

\noindent{\bf Acknowledgement.} This work was supported in part by
AFOSR Grant \# FA9550-09-1-0165 and ARO Grant \# W911NF-13-1-0370.


\begin{thebibliography}{10}


{\small

\bibitem{SchneiderIPL1985}
B.~Alpern and F.~Schneider.
\newblock Defining liveness.
\newblock {\em Inf. Process. Lett.}, 21(4):181--185, 1985.

\bibitem{CanettiFOCS01}
R.~Canetti.
\newblock Universally composable security: A new paradigm for cryptographic
  protocols.
\newblock In {\em Proc. FOCS'01}, pp 136--145.

\bibitem{CanettiEurocrypt03}
R.~Canetti, E.~Kushilevitz, and Y.~Lindell.
\newblock On the limitations of universally composable two-party computation
  without set-up assumptions.
\newblock In EUROCRYPT'03, pp 68--86.

\bibitem{WangTISSEC08}
D.~Chakrabarti, Y.~Wang, C.~Wang, J.~Leskovec, and C.~Faloutsos.
\newblock Epidemic thresholds in real networks.
\newblock {\em ACM Trans. Inf. Syst. Secur.}, 10(4):1--26, 2008.

\bibitem{ClarksonJCS2010}
M.~Clarkson and F.~Schneider.
\newblock Hyperproperties.
\newblock {\em Journal of Computer Security}, 18(6):1157--1210, 2010.

\bibitem{ErdiBook2008}
P.~Erdi.
\newblock {\em Complexity Explained}.
\newblock Springer, 2008.

\bibitem{GligorSecurityProtocol04}
V.~Gligor.
\newblock Security of emergent properties in ad-hoc networks.
\newblock In {\em Proc. Security Protocols Workshop'04}, pp 256--266.

\bibitem{GoguenOakland82}
J.~Goguen and J.~Meseguer.
\newblock Security policies and security models.
\newblock In {\em IEEE Symp. on Security \& Privacy'82}, pp 11--20.

\bibitem{GMW87}
O.~Goldreich, S.~Micali, and A.~Wigderson.
\newblock How to play any mental game or a completeness theorem for protocols
  with honest majority.
\newblock In {\em Proc. ACM STOC'87}, pp  218--229.

\bibitem{HintonNSPW97}
H.~Hinton.
\newblock Under-specification, composition and emergent properties.
\newblock In {\em Proc. NSPW'97}, pp. 83--93.

\bibitem{Kubik:2002:TFE:778855.778861}
A.~Kub\'{\i}k.
\newblock Toward a formalization of emergence.
\newblock {\em Artif. Life}, 9(1):41--65, 2002.

\bibitem{LamportIEEETSE77}
L.~Lamport.
\newblock Proving the correctness of multiprocess programs.
\newblock {\em IEEE Trans. Software Eng.}, 3(2):125--143, 1977.

\bibitem{LindellFOCS03}
Y.~Lindell.
\newblock General composition and universal composability in secure multi-party
  computation.
\newblock In {\em FOCS'03}, pp 394--403.

\bibitem{SchneiderDHC2011}
F.~Schneider.
\newblock Beyond traces and independence.
\newblock In {\em Dependable and Historic Computing}, pp 479--485, 2011.

\bibitem{XuHotSOS14-cybersecurity-dynamics-poster}
S.~Xu.
\newblock Cybersecurity dynamics.
\newblock In {\em HotSOS'14 (poster)}.

\bibitem{XuInternetMath2013-sub}
S.~Xu, W.~Lu, and H.~Li.
\newblock A stochastic model of active cyber defense dynamics.
\newblock {\em Internet Mathematics}, 2014 (to appear).

\bibitem{XuTAAS2012}
S.~Xu, W.~Lu, and L.~Xu.
\newblock Push- and pull-based epidemic spreading in arbitrary networks:
  Thresholds and deeper insights.
\newblock {\em ACM TAAS},  7(3):32:1--32:26, 2012.

\bibitem{XuTAAS2014}
S. Xu, W. Lu, L. Xu, and Z. Zhan. 
\newblock Adaptive Epidemic Dynamics in Networks: Thresholds and Control. 
\newblock {\em ACM TAAS}, 8(4):19 (2014)

\bibitem{XuTDSC2012}
S.~Xu, W.~Lu, and Z.~Zhan.
\newblock A stochastic model of multivirus dynamics.
\newblock {\em IEEE TDSC},
  9(1):30--45, 2012.


\bibitem{Yao86}
A.~C. Yao.
\newblock How to generate and exchange secrets.
\newblock In {\em Proc.\ FOCS'86},  pp 162--167.

}

\end{thebibliography}
\end{document}